\documentclass[conference]{IEEEtran}
\IEEEoverridecommandlockouts

\usepackage{cite}
\usepackage{amsmath,amssymb,amsfonts,mathrsfs,amsthm}
\usepackage{algorithm}
\usepackage{algpseudocode}
\usepackage{graphicx}
\usepackage{textcomp}
\usepackage{xcolor}
\usepackage{url}
\usepackage{adjustbox}
\usepackage{balance}
\usepackage{comment}
\usepackage{makecell}

\newtheorem{remark}{Remark}

\begin{document}

\title{Mobile Base Station Optimal Tour in \\Wide Area IoT Sensor Networks}

\author{\IEEEauthorblockN{Sachin Kadam}
\thanks{This research was partially supported by the Dean R\&C, MNNIT Allahabad, India, through the SEED grant (334/R\&C/2025-26), and also by the Anusandhan National Research Foundation (ANRF), Government of India, under the Prime Minister Early Career Research Grant (PM-ECRG) with File Number: \text{ANRF/ECRG/2025/003076/ENS}.}
\IEEEauthorblockA{{Department of Electronics and Communication Engineering} \\
{Motilal Nehru National Institute of Technology (MNNIT) Allahabad, Prayagraj, UP 211004, India}\\
Email: sachink@mnnit.ac.in}
}

\maketitle

\begin{abstract}
Wide-area IoT sensor networks require efficient data collection mechanisms when sensors are dispersed over large regions with limited communication infrastructure. Unmanned aerial vehicle (UAV)-mounted Mobile Base Stations (MBSs) provide a flexible solution; however, their limited onboard energy and the strict energy budgets of sensors necessitate carefully optimized tour planning. In this paper, we introduce the Mobile Base Station Optimal Tour (MOT) problem, which seeks a minimum-cost, non-revisiting tour over a subset of candidate stops such that the union of their coverage regions ensures complete sensor data collection under a global sensor energy constraint. The tour also avoids restricted areas. We formally model the MOT problem as a combinatorial optimization problem, which is NP-hard. Owing to its computational intractability, we develop a polynomial-time greedy heuristic that considers minimizing MBS travel cost covering all IoT sensors while avoiding restricted areas. Using simulations, we obtain tours with low cost, complete sensor coverage, and faster execution. 
The proposed framework provides both theoretical insight into the structural complexity of MBS-assisted data collection and a practical algorithmic solution for large-scale IoT deployments.
\end{abstract}

\section{Introduction}\label{Sec:Introduction}
The convergence of the Internet of Things (IoT) and Wireless Sensor Networks (WSNs) plays a vital role in advancing agricultural and industrial productivity. These technologies support a wide range of farming operations, including intelligent irrigation, soil moisture sensing, optimized fertilizer application, early detection of pests and crop diseases, and improved energy management~\cite{mowla2023internet}. Large-scale IoT sensor (IoT node interfaced with a sensor) deployments are also becoming increasingly common in environmental surveillance, precision agriculture, and disaster response applications~\cite{mozaffari2019tutorial,zeng2019cellular}. In areas lacking reliable communication infrastructure, Mobile Base Stations (MBSs), especially those mounted on unmanned aerial vehicles (UAVs), offer a practical and rapidly deployable solution for collecting sensor data~\cite{lin2018sky}. Nevertheless, the limited onboard energy of UAV platforms and communication constraints necessitate careful trajectory optimization to minimize travel costs and prolong mission duration~\cite{wang2024trajectory}. Furthermore, IoT sensors typically operate under strict energy budgets, making it essential to control transmission power during data upload to ensure long-term network sustainability~\cite{cai2024joint}. In practical deployments, an MBS often visits selected candidate stops while avoiding restricted areas\footnote{These areas include no-fly zones or regions surrounding military facilities.}, whose combined coverage must include all sensors. This motivates the Mobile Base Station Optimal Tour (MOT) problem, which aims to determine a minimum-cost, non-revisiting route that guarantees complete data collection while adhering to sensor energy constraints and avoiding restricted areas. The MOT formulation leads to a combinatorial optimization problem distinct from conventional continuous trajectory design approaches.
\begin{figure}
    \centering
    \includegraphics[width=0.975\linewidth]{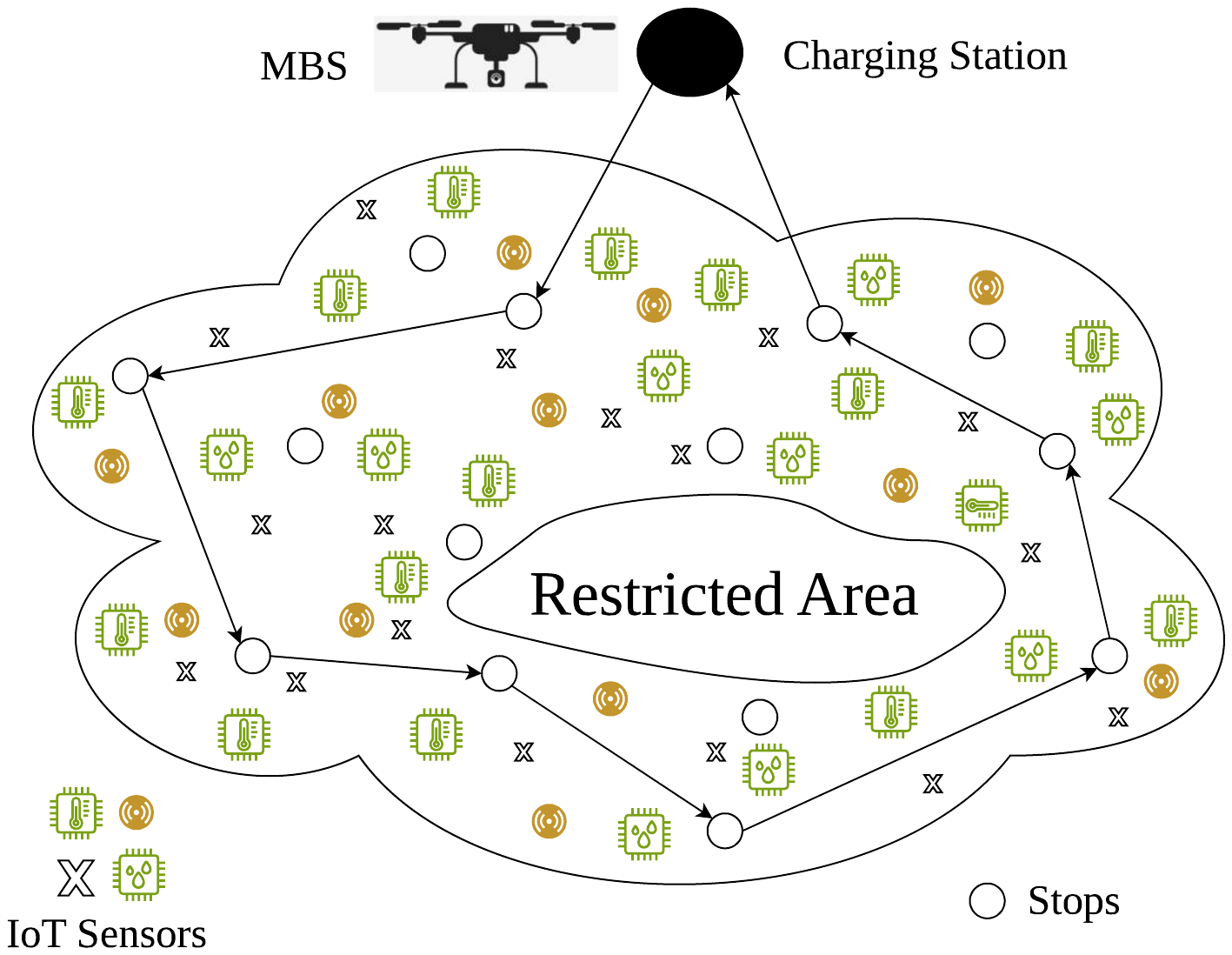}
    \caption{Illustration of a mobile base station (MBS) tour through a wide area region while avoiding restricted areas and stopping at multiple predetermined locations such that the combined coverage of these stops includes all deployed IoT sensors.
}
    \label{fig:SystemModel}
\end{figure}

We consider an MBS that traverses a region, avoiding the forbidden areas, and stops at multiple predetermined locations such that the combined coverage areas of these stops include all the deployed IoT sensors (see Fig.~\ref{fig:SystemModel}). The MBS is battery-powered and begins its mission from a charging station, which also serves as its starting and ending point. It departs from this station, visits selected stops to collect data from the IoT sensors inside its coverage area, and finally returns to the origin. Due to its limited energy supply, the travel route must be carefully optimized to reduce power consumption. Assume a set of candidate stops is given. Moving between any two stops incurs a travel cost. The energy usage minimization of IoT sensors  is essential since they operate with very limited energy resources. From the available candidate stops, the MBS must select a subset and determine a tour, while avoiding the forbidden areas, such that each IoT sensor is covered at least once while each stop is visited no more than once. Based on these considerations, we define the MBS Optimal Tour (MOT) problem. 
{The objective of the MOT problem is to determine an MBS tour with minimum travel cost while satisfying the following constraints: (i) the total energy consumed by the IoT sensors during data transmission does not exceed the prescribed threshold; (ii) all deployed IoT sensors are successfully served during the tour; (iii) the MBS avoids restricted areas; and (iv) each candidate stop is visited at most once.
}
We formulate the MOT problem as an optimization model, and since it is an NP-hard problem, we also develop a greedy algorithm for its solution.

\section{Related Work}\label{Sec:RelatedWork}
Recent studies have explored UAV-assisted data collection and estimation of active IoT sensors in wide-area IoT networks, primarily focusing on trajectory optimization and energy efficiency~\cite{baek2020optimal,li2023data,baek2020energy,zhu2023flying,kadam2020spcom,KADAM2024103779}. Multi-objective routing frameworks such as~\cite{wang2024trajectory} jointly optimize coverage, delay, and energy consumption. Comprehensive surveys such as \cite{comcom2023_uav_survey} review UAV deployment strategies, trajectory optimization, and aerial base station design, highlighting challenges related to mobility management, coverage planning, and energy efficiency. These works provide a broad foundation but do not explicitly address discrete stop selection and constrained tour construction. Heuristic placement and trajectory optimization techniques have also been proposed for smart farming and emergency scenarios~\cite{huang2024drl,liu2024multi,eldeeb2023traffic}. These works emphasize flight path shaping and resource scheduling but generally overlook combinatorial tour construction over a predefined set of candidate stops with guaranteed full coverage. In contrast, the proposed MOT formulation explicitly integrates discrete stop selection, non-revisiting tour construction, complete sensor coverage, and a hard constraint on total IoT sensor energy expenditure. This leads to a fundamentally different combinatorial optimization problem, which is NP-hard, distinguishing it from existing continuous trajectory and learning-based approaches. In a static WSN, the study in~\cite{PSO_Na} employs particle swarm optimization (PSO) to determine the optimal positions for UAVs, ensuring full network connectivity and maximizing the total collected sensor data. This approach uses multiple UAVs, which form a UAV network, whereas in our work we consider a single UAV. 

Efficient data gathering in UAV-enabled IoT networks has also received significant attention. The survey work in~\cite{joshi2023_uav_iot_survey} discusses reliability and energy-aware data collection mechanisms, while the study in~\cite{search7_2024_uav_sd} investigates UAV-enabled adaptive data collection in software-defined WSNs. Although these works consider path design and network adaptation, they primarily focus on continuous trajectory optimization rather than combinatorial tour formulation. In contrast, the proposed MOT problem formulates data collection as a discrete combinatorial optimization problem that jointly considers stop selection, non-revisiting tour construction, restricted area avoidance, complete sensor coverage, and constrained IoT sensor energy expenditure, leading to a formulation distinct from prior continuous trajectory-based approaches.

{The proposed MOT formulation is related to classical combinatorial routing problems such as the Covering Salesman Problem and the Generalized Traveling Salesman Problem~\cite{kleinberg2006algorithm}. However, unlike these formulations, the MOT problem explicitly incorporates aggregate IoT sensor energy constraints derived from the underlying communication process together with restricted-area avoidance and MBS-assisted sensor coverage requirements. These additional considerations lead to a problem formulation tailored to wide-area IoT data collection scenarios.
}

\section{System Model}\label{Sec:System_Model}
During the exchange of sensor information between IoT devices and the MBS, packet losses may arise due to variations in the reliability of the connection between each device and the MBS. In this study, link reliability is characterized by the probability of successful packet delivery. This reliability is influenced by the properties and conditions of the communication channel connecting the transmitter and the receiver.

In~\cite{mahmood2016packet}, the average packet error rate $\bar{\eta}(\bar{\gamma})$ over a Rayleigh fading channel is expressed as
\begin{equation}
\bar{\eta}(\bar{\gamma})\! \approx \!1 \!- e^{-\frac{a_n}{\bar{\gamma}}} \Gamma\!\left(\!1+\frac{b_n}{\bar{\gamma}}\!\right)\!,
a_n \!= \!\frac{\log (nc_m)}{k_m}, b_n \!= \!\frac{1}{k_m},
\label{eq:exponential_aNbN}
\end{equation}
where $\bar{\gamma}$ is the average SNR, $\Gamma(\cdot)$ is the gamma function, $n$ is the packet length in bits, and $c_m$ and $k_m$ are modulation-dependent constants.\footnote{For example, for FSK, $c_m = 1/2$ and $k_m=1/2$ ; for BPSK $c_m = 1$ and $k_m=2$.} In this work, we assume that the average SNR between the IoT devices and the MBS can be determined in advance. This average SNR primarily depends on the separation distance, and it can be evaluated because the stopping points of the MBS are predetermined. Let $\mathcal{S}$ represent the collection of IoT sensors and $\mathcal{T}$ denote the set of MBS stops. Within an MBS-assisted wireless system, any unsuccessful packet delivery triggers a retransmission process. Let  the maximum transmission counter for a packet be $q_{max}$, then the average packet transmission success rate, $\bar{\rho}_{s,\tau}$, over the link between the sensor $s \in \mathcal{S}$ and the MBS when it is at stop $\tau \in \mathcal{T}$ is computed as,
\begin{equation}
\bar{\rho}_{s,\tau} = 1 - 
[1-\bar{\eta}(\bar{\gamma}_{s,\tau})]^{q_{max}} ,
\label{eq:PSU_rayA}
\end{equation}
where $\bar{\gamma}_{s,\tau}$ is the average SNR between the IoT sensor $s$ and the MBS when it is at stop $\tau$. The expected number of retransmissions before the successful delivery of a data packet is
\begin{equation}
    R_{s,\tau} = \frac{\bar{\rho}_{s,\tau}}{\bar{\eta}(\bar{\gamma}_{s,\tau})}.
\label{eq:E_PSU_rayA}    
\end{equation}

We define a threshold packet delivery probability, denoted by $\bar{\rho}$, which guarantees that the MBS can reliably collect data from any IoT device. Accordingly, the requirement $\bar{\rho}_{s,\tau} \ge \bar{\rho}$ guarantees successful data reception from sensor $s \in \mathcal{S}$ when the MBS is positioned at stop $\tau \in \mathcal{T}$.

At stop $\tau$, the MBS coverage area comprises the ground region where IoT devices can directly communicate with it. To collect sensor data, the MBS operates at a fixed minimum altitude $h_{min}$, which may be required to maintain reliable communication given the limited energy of the IoT devices.

As data is sent from an IoT sensor to the MBS, the signal experiences attenuation caused by path loss. Under the assumption of free-space propagation (path loss exponent equal to 2) and neglecting fading effects, the received signal power at the MBS when it is located at stop $\tau$, denoted by $P_{\tau}^{r}$, can be formulated under ideal conditions as follows:\footnote{Here, $P^{t}_{s}$ denotes the transmission power of sensor $s$, $G_{s}^{t}$ is the sensor’s transmit antenna gain, $G_{\tau}^{r}$ represents the MBS’s receive antenna gain at stop $\tau$, $\lambda_{s}$ is the wavelength of the signal sent by sensor $s$, and $d_{s,\tau}$ is the distance between the sensor and the MBS. {Although free-space propagation provides a tractable analytical model, more sophisticated propagation models may be incorporated within the proposed framework.}}
\begin{equation}
P_{\tau}^{r} =
G_{s}^{t} \cdot G_{\tau}^{r} \cdot \frac{\lambda_{s}^{2}}{(4\pi)^{2}}
\left(\frac{1}{d_{s,\tau}}\right)^{2}
P^{t}_{s}.
\end{equation}
For reliable wireless reception, the received signal strength must be at least equal to the minimum detectable threshold, denoted by $P_{\tau,\min}^{r}$. Accordingly, the greatest allowable separation distance, $d_\tau^{\max}$, at which the MBS positioned at stop $\tau$ can successfully collect data from sensor $s$, is determined as follows:
\begin{equation}
d_\tau^{\max} =
\frac{\lambda_{s}}{4\pi}\sqrt{
\frac{
G_{s}^{t} \cdot G_{\tau}^{r} \cdot  P_{s}^{t}
}{ P_{\tau,\min}^{r}}
}.
\end{equation}

Define the binary variable $Z_{s,\tau} \in \{0,1\}$ such that it equals 1 if sensor $s$ lies within the MBS coverage area when the MBS is positioned at stop $\tau$, meaning that $d_{s,\tau} \le d_{\tau}^{\max}$; otherwise, it is set to 0.\footnote{{The binary quantity $Z_{s,\tau}$ is a precomputed coverage parameter determined by the geometry of the network deployment. Since the candidate MBS stops are assumed to be predetermined, the distance $d_{s,\tau}$ between sensor $s$ and stop $\tau$ is known a priori. Consequently, the value of $Z_{s,\tau}$ can be computed before solving the MOT problem
and does not constitute a decision variable of the optimization problem.
}} Accordingly, $Z_{s,\tau}$, for all $s \in S$ and $\tau \in T$, is defined as
\begin{equation}
Z_{s,\tau}:=
\begin{cases}
1, & d_{s,\tau}\leq d^{\max}_{\tau},\\
0, & \text{otherwise}.
\end{cases}
\end{equation}
Similarly, define the binary variable $D_{s,\tau} \in \{0,1\}$ such that it equals 1 whenever the sensor $s$ has data to send to the MBS when it is at stop $\tau$. Once the data is transferred to the MBS when it is at stop $\tau_i$, $D_{s,\tau_j} = 0$ for $j=i+1, \ldots, |\mathcal{T}|$. This condition is met whenever $\bar{\rho}_{s,\tau_i} \ge \bar{\rho}_{min}$ for all $s \in \mathcal{S}$ and $i = 1,\ldots,|\mathcal{T}|$.

Similarly, let $P^t_{s,\tau} \in~\{0,P^t_{s}\}$ be the transmission power of the sensor $s$ when the MBS is at the stop $\tau$ and it is defined as follows:
\begin{align}
    P^t_{s,\tau} := 
    \begin{cases}
    P^t_{s}, & d_{s,\tau} \le d_\tau^{\max} \text{ AND } D_{s,\tau} = 1 \\
    0, &\mbox{otherwise}.
    \end{cases}
\end{align}
So, $P^t_{s,\tau} = P^t_{s} Z_{s,\tau} D_{s,\tau}$.

\subsection{MBS Optimal Tour (MOT) Problem}\label{MOT}
Consider a connected directed graph $\mathcal{G} = (\mathcal{T}, \mathcal{E})$, where $\mathcal{T}$ is the set of candidate stops for the MBS in the target area, and $\mathcal{E}$ is the set of directed edges linking these stops. The sets have cardinalities $|\mathcal{T}| = M+1$ and $|\mathcal{E}| = E$. Let $\mathcal{S}$ represent the set of IoT sensors deployed in the region, with $|\mathcal{S}| = N$. The travel cost for moving from stop $u$ to stop $v$ along edge $(u,v)$ is denoted by $c_{u,v}$.
 
This work aims to determine the MBS’s optimal route during its operational period to minimize the overall travel cost. The MBS departs from the charging station, designated as stop $0 \in \mathcal{T},$\footnote{It is assumed that no sensors fall within the MBS coverage area while it is located at the charging station; hence, $Z_{s,0} = 0$ for all $s \in \mathcal{S}$.} then proceeds to selected stops to gather data from the IoT sensors. After completing data collection from all the sensors, the MBS returns to stop $0$. Each stop, except for the charging station, is visited at most once during a tour.

Let $\mathcal{T}_o \subseteq \mathcal{T}$ denote the subset of stops included in a particular MBS tour, where $|\mathcal{T}_o| = T_o + 1 \le M + 1$. For $i \in {0,1,\ldots,T_o}$, let $\tau_i \in \mathcal{T}_o$ represent the $i$-th stop visited along the route, with $\tau_0 = \tau_{T_o+1} = 0$. The tour can therefore be described by the ordered sequence of edges $(\tau_i, \tau_{i+1})$ for $i \in {0,1,\ldots,T_o}$.
To conserve sensor energy, each IoT sensor is switched off after successfully transmitting its data to the MBS and remains inactive until the next operational cycle. Specifically, for a given sensor $s$, the transmit power satisfies $P^t_{s,\tau_j} = 0$ for $j = i+1,\ldots,T_o$ once a successful transmission occurs at stop $\tau_i$. This condition is met whenever $\bar{\rho}_{s,\tau_i} \ge \bar{\rho}_{min}$ for all $s \in \mathcal{S}$ and $i = 1,\ldots,T_o$.

The optimization task, referred to as the MOT problem, aims to determine a route for the MBS, avoiding restricted area $\mathcal{F}$, such that each sensor in $\mathcal{S}$ is covered by at least one of the MBS positions corresponding to the stops in $\mathcal{T}_o$. The objective is to minimize the total travel cost while keeping the sensor network’s energy consumption within the prescribed limit $\overline{P}_{max}$. This problem can be formulated mathematically as follows:
\begin{subequations}
\begin{align} 
\min_{\substack{T_o,~\tau_i, \\ i \in \{1,\ldots, T_o\}}} & \sum_{i=0}^{T_o} c_{\tau_i,\tau_{i+1}}   \label{eq:obj_fun}\\
\textrm{s.t.} \quad & \sum_{i=1}^{T_o} \sum_{s \in \mathcal{S}} P^t_{s,\tau_i} R_{s,\tau_i}  \le \overline{P}_{max}, \label{eq:power_constraint}\\ 
\quad & {\sum_{i=1}^{T_o}
Z_{s,\tau_i}D_{s,\tau_i}
\geq 1,
\qquad \forall s\in S,} \label{eq:coverage_constraint}\\
\quad & (\tau_i,\tau_j) \notin \mathcal{F}, \label{eq:RestrictedArea_constraint} \\
\quad & \tau_0 = \tau_{T_o+1} = 0, \label{eq:BeginEndStops_constraint} \\
\quad & \tau_i \neq \tau_j, \,  \forall i,j \in \{1, \ldots, T_o\}, \, i \neq j, \label{eq:UniqueStop_constraint}\\
\quad & T_o \in \{1, \ldots, M\}, \label{eq:Mstar_constraint}\\
\quad & D_{s,\tau} \in\{0, 1\},~ \overline{P}_{max}, P^t_{s,\tau}, R_{s,\tau}, c_{\tau_i,\tau_j} \in \mathcal{R},  \nonumber\\
\quad & \quad \forall \tau \in \{0,1, \ldots, M\}, \forall s \in \mathcal{S}. \nonumber
\end{align}
\end{subequations}
The goal of the MOT problem, as expressed in~\eqref{eq:obj_fun}, is to minimize the total travel cost incurred by the MBS.
{The energy constraint in~\eqref{eq:power_constraint} is imposed at the network level and limits the aggregate energy expenditure of all sensors during a data collection cycle.}\footnote{{Modeling individual sensor energy budgets constitutes an interesting extension and is left for future investigation.}
}
{The condition in~\eqref{eq:coverage_constraint} guarantees that every IoT sensor is served at least once during the MBS tour. 
Therefore, complete data collection from all deployed sensors is ensured.
}
The condition in~\eqref{eq:RestrictedArea_constraint} ensures that no path of the MBS tour passes through the restricted area $\mathcal{F}$. The constraints in~\eqref{eq:BeginEndStops_constraint} and~\eqref{eq:UniqueStop_constraint} enforce that the MBS starts and ends its tour at the charging station and that every other stop in $\mathcal{T}_o$ is visited at most once, respectively.
Finally, the feasible range of values for $T_o$ is defined in~\eqref{eq:Mstar_constraint}.

{The MOT problem is NP-hard. The classical traveling salesman problem (TSP) can be viewed as a special case of the MOT problem obtained when every stop must be visited, all sensors are covered with the stops, the energy constraint is inactive, and no restricted areas are present. Since TSP is NP-hard~\cite{kleinberg2006algorithm}, MOT problem is NP-hard. The decision version of the MOT problem is NP-complete; the proof is left to future studies.}
To tackle it, a greedy algorithm is proposed, as detailed in the following subsection.

\subsection{Greedy Algorithm} \label{Section:Greedy}
We assume that the graph $\mathcal{G}$ is complete, i.e., there exists a directed link between every pair of stops. Let $\mathcal{K}_{\tau_i}$ denote the set of IoT sensors covered by the MBS up to stop $\tau_i$, for $i \in \{0,1,\ldots,T_o\}$, and let $k_{\tau}$ denote the set of sensors covered specifically at stop $\tau$. Define
\begin{equation}
\overline{\zeta}_{\tau_i} = \sum_{s \in \mathcal{S}} P^t_{s,\tau_i} R_{s,\tau_i}, 
\quad i \in \{1,\ldots,T_o\},
\end{equation}
which represents the cumulative energy consumption of the sensors up to stop $\tau_i$. Equivalently, $\overline{\zeta}_{\tau_i}$ can be written recursively as
\begin{equation}
\overline{\zeta}_{\tau_i} 
= \overline{\zeta}_{\tau_{i-1}} 
+ \sum_{j \in k_{\tau_i}} \zeta_j,
\end{equation}
with the initialization $\overline{\zeta}_{\tau_0} = 0$. 
The assigned cost value between stops $u$ and $v$, if the path ($u,v$) goes via a forbidden area is $\infty$.
Let $\mathcal{U}_i$ denote the set of stops not yet visited by the MBS after reaching stop $\tau_i$, for $i \in {0,1,\ldots,T_o}$. The MOT problem aims to construct a tour with minimal travel cost. Using a greedy approach, the MBS begins at the charging station $\tau_0$ and iteratively selects the next stop $\tau \in \mathcal{U}_i$ with the lowest travel cost $c_{\tau_i,\tau}$,\footnote{{For instance, the travel cost can be modeled as inversely proportional to the uncovered sensor gain associated with a stop, such that stops offering higher uncovered gain incur lower travel costs.}} collecting data from sensors within its coverage. After each stop, the MBS checks whether all sensors in $\mathcal{S}$ have been covered; if so, it returns to $\tau_0$, completing the tour. Otherwise, it evaluates the cumulative sensor energy consumption. If the total exceeds $\overline{P}_{max}$, the tour ends at the current stop. If the energy constraint is still satisfied, the process continues from the current location. The complete greedy procedure is summarized in Algorithm~\ref{alg:greedy}.

It is worth noting that the greedy algorithm performs particularly well when the power constraint $\overline{P}_{max}$ is sufficiently large. In such cases, it is unlikely that the cumulative sensor energy consumption exceeds $\overline{P}_{max}$ before all sensors are covered.

\begin{algorithm}
\caption{Greedy Algorithm}\label{alg:greedy}
\begin{algorithmic}
\State \textbf{Input:} $\mathcal{T}$, $\mathcal{F}$,  $\overline{P}_{max}$, $c_{u,v},$ $u,v \in \mathcal{S}$, $u \neq v$, $\overline{\gamma}$, $N, c_m, k_m$, $q_{max}$, $d_\tau^{max}$, $d_{s,\tau}$, $s \in \mathcal{S}$, $\tau \in \mathcal{T}$
\State Initialize $i=0$, $\tau_0 = 0$, $\mathcal{T}_o \!\gets \{0\}$, $\mathcal{K}_{\tau_0} \!\!\gets \emptyset$, $\overline{\zeta}_{\tau_0} \!\!\gets 0$, $\mathcal{U}_0\!\! \gets \mathcal{T} \backslash \{0\}$, $D_{s,0} = 1, \forall s \in \mathcal{S}, ~c_{u,v} = \infty \text{ for } (u,v) \in \mathcal{F}$
\While {$\mathcal{K}_{\tau_i} \neq \mathcal{T}$ AND $\overline{\zeta}_{\tau_i} \le \overline{P}_{max}$}
\State Choose stop $\tau \in \mathcal{U}_i$ such that $c_{\tau_i,\tau}$ is smallest. 
\State $\mathcal{K}_{\tau_{i+1}} \gets \mathcal{K}_{\tau_i} \bigcup k_{\tau}$
\State $\overline{\zeta}_{\tau_{i+1}} \gets \overline{\zeta}_{\tau_i} + \sum_{j \in k_{\tau}}\zeta_{j}$
\State $\mathcal{U}_{i+1} \gets \mathcal{U}_i \setminus \{\tau\}$
\State $\tau_{i+1} \gets \tau$; $\mathcal{T}_o \gets \mathcal{T}_o \bigcup \{\tau\}$
\State $i \gets i+1$
\EndWhile
\State $\tau_{i+1} \gets 0$
\State $T_o \gets |\mathcal{T}_o| - 1$
\State \textbf{Output:} $\mathcal{T}_o$, $(\tau_i, \tau_{i+1}), i=0, \ldots, T_o$.
\end{algorithmic}
\end{algorithm}

\begin{remark}[Computational Complexity]
The proposed greedy algorithm for the MOT problem runs in
$
\mathcal{O}(M^2 + MN)
$
time and requires $\mathcal{O}(M + N)$ space.
\end{remark}
An exact solution to the MOT problem requires enumerating all feasible non-revisiting tours over subsets of candidate stops. 
In the worst case, when all $M$ stops must be considered, the number of possible tours grows factorially as $\mathcal{O}(M!)$. 
Hence, a brute-force optimal search incurs exponential time complexity.
Compared to the exponential complexity of exhaustive search, the proposed greedy algorithm provides a polynomial-time alternative with complexity $\mathcal{O}(M^2 + MN)$, making it computationally scalable for large-scale IoT sensor networks.

{The proposed greedy algorithm emphasizes computational efficiency and scalability for large-scale IoT deployments. Although metaheuristic approaches, such as genetic algorithms and PSO algorithms, may achieve improved solution quality, they generally incur substantially higher computational overhead. The development of such approaches for the MOT problem remains an interesting direction for future work.
}
\section{Simulation Results}\label{Sec:Results}
Simulations are performed using MATLAB version R2025b. The host machine has an AMD Ryzen 9 9950X 16-Core processor  with 64 GB RAM clocking at 4.30 GHz. The proposed greedy algorithm is evaluated on a simulated WSN consisting of $N=100$ IoT sensors,   deployed using Poisson disk sampling,\footnote{Poisson disk sampling is a method for generating randomly distributed points while ensuring that each point is separated by at least a minimum distance of $d_{min}=8\text{ m}$.} in a $100\times100$~m$^2$ area, with a single restricted zone of $20\times20$~m$^2$. 
The MBS visits the stop that covers the highest number of previously uncovered sensors at each iteration, choosing stops based on a minimum cost—in this case, the maximum uncovered sensor gain.  
In order to prevent the MBS from passing through restricted areas, candidate stops whose paths cross them are skipped. When every IoT sensor is covered, the simulation ends the tour, and the MBS forms a closed-loop trajectory by returning to its initial starting point.\footnote{{During the simulation, the aggregate sensor energy expenditure, computed according to~\eqref{eq:power_constraint}, was monitored throughout the MBS tour. In the considered deployment scenario, the proposed algorithm successfully satisfied the prescribed energy constraint while achieving complete sensor coverage. This observation demonstrates the ability of the MOT framework to jointly account for routing efficiency and energy-aware data collection.
}}

\begin{figure}[ht!]
    \centering
    \includegraphics[width=0.48\textwidth]{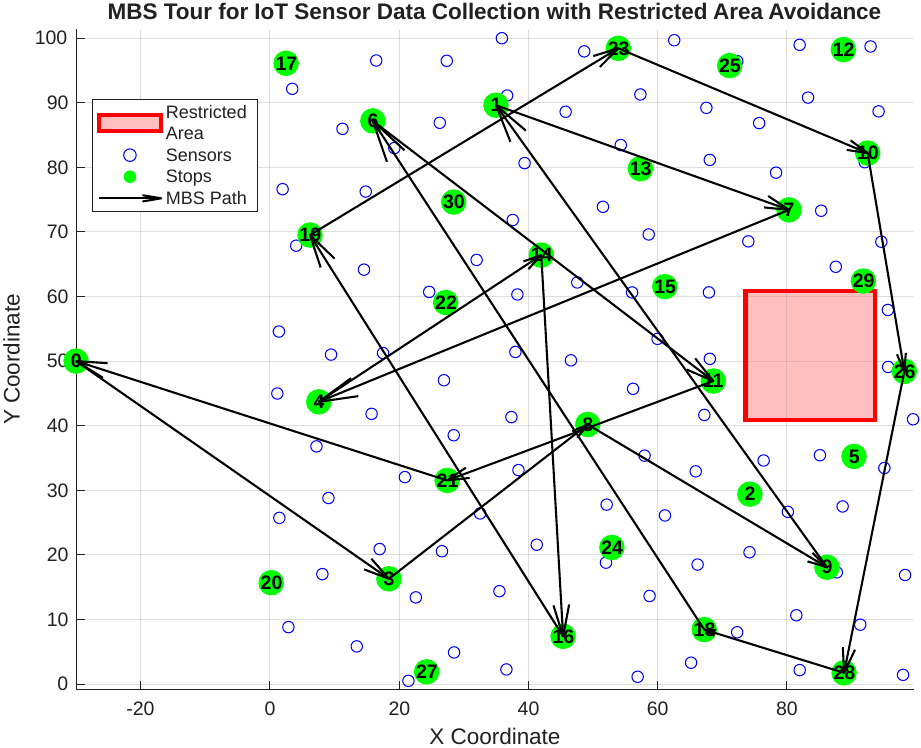} 
    \caption{MBS data collection tour in a wireless IoT sensor network with a restricted area. Blue hollow circles represent IoT sensor nodes, green filled circles indicate selected MBS stops with numbers, black lines with arrows indicate the MBS tour path, and the red shaded region shows the restricted area. The MBS returns to its starting stop to form a closed-loop trajectory. This tour achieves full sensor coverage while minimizing the total cost of the tour with a constraint on the total energy of the network and restricted area avoidance. The MBS tour $0 \rightarrow 3 \rightarrow 8 \rightarrow 9 \rightarrow 1 \rightarrow 7 \rightarrow 4 \rightarrow 14 \rightarrow 16 \rightarrow 19 \rightarrow 23 \rightarrow 10 \rightarrow 26 \rightarrow 28 \rightarrow 18 \rightarrow 6 \rightarrow 11 \rightarrow 21 \rightarrow 0$ has a length of approximately 178 m, and the algorithm execution time is 0.12 s.}
    \label{fig:mbs_tour}
\end{figure}

The simulation results demonstrate the efficiency of the proposed greedy approach. In a typical run, 100\% of sensors are successfully covered by the selected MBS stops. In the shown example of Fig.~\ref{fig:mbs_tour}, the MBS visited only 17 stops out of 30 possible candidate stops before completing the full coverage, indicating a significant reduction in unnecessary visits. The MBS tour obtained as shown in Fig.~\ref{fig:mbs_tour}, has a length, calculated as the sum of Euclidean distances between consecutive stops, of approximately 178 m, and the algorithm execution time is just 0.12 s, including the return to the charging station.\footnote{When results are averaged over 50 independent sensor deployments generated using Poisson disk sampling, the results are 178 $\pm$ 12 m and 0.12 $\pm$ 0.01 s.}
These results highlight the ability of the algorithm to complete the  coverage with minimal time and cost. Figure~\ref{fig:mbs_tour} illustrates the sensor locations, selected stops, MBS path, and restricted area, providing an intuitive representation of the tour with forbidden area avoidance.
\section{Conclusions and Future Work}\label{Sec:Conclusions}
This paper studied the Mobile Base Station Optimal Tour (MOT) problem for wide-area IoT sensor networks, where a UAV-mounted MBS must construct a minimum-cost, non-revisiting tour that guarantees full sensor coverage under a global energy constraint while avoiding the restricted areas. We formulated the problem as a combinatorial optimization model, which is an NP-hard problem, demonstrating its computational intractability. To obtain feasible solutions, we proposed a polynomial-time greedy heuristic that minimizes travel cost while respecting IoT sensor energy limits and avoids restricted areas. Simulation results confirmed the effectiveness of the approach in achieving the low-cost tour and faster execution of the algorithm. 
{Since the proposed greedy algorithm relies on local decisions, it does not guarantee globally optimal solutions and may exhibit degraded performance in scenarios where locally attractive stop selections lead to inefficient tours. Future work will focus on designing approximation algorithms with provable bounds, investigating metaheuristic approaches, extending the framework to multi-UAV settings, and incorporating stochastic energy models.
} 

\bibliographystyle{ieeetr}
\bibliography{references.bib}
\end{document}